\documentclass[11pt,twoside]{article}

\usepackage{asp2004}
\usepackage{epsf}
\usepackage{lscape}
\begin{document}
\title{Density Distribution of the Warm Ionized Medium}
\author{Alex S. Hill\footnotemark[1], Ronald J. Reynolds\footnotemark[1], Robert A. Benjamin\footnotemark[2], and L. Matthew Haffner\footnotemark[1]}
\affil{\footnotemark[1]University of Wisconsin-Madison (hill@astro.wisc.edu), \footnotemark[2]University of Wisconsin-Whitewater}

\begin{abstract}
Observations of H$\alpha$ emission measures and pulsar dispersion measures at high Galactic latitude ($|b| > 10\deg$) provide information about the density and distribution of the diffuse warm ionized medium (WIM). The diffuse WIM has a lognormal distribution of $\textrm{EM} \sin |b|$, which is consistent with a density structure established by isothermal turbulence. The H$^+$ responsible for most of the emission along high-$\textrm{EM} \sin |b|$ sightlines is clumped in high density ($> 0.1 \textrm{ cm}^{-3}$) regions that occupy only a few parsecs along the line of sight, while the H$^+$ along low-EM sightlines occupies hundreds of parsecs with considerably lower densities.
\end{abstract}

\section{Introduction}

The warm ionized medium (WIM) is a major component of our Galaxy. The WIM consists of a pervasive, diffuse plasma layer with a temperature $T \sim 8000$~K, a scale height of $h \approx 1$~kpc (more than $3$ times the scale height of neutral hydrogen), and densities of $\approx 0.08 \textrm{ cm}^{-3}$ which occupy $\approx 20\%$ of the volume of the disk \citep{r91}. The ionization and heating sources of the WIM and its relationship to the other phases of the ISM are not well understood.

We use observations of interstellar H$\alpha$ emission and pulsar dispersion measures to explore the density and distribution of diffuse plasma in the WIM. Pulsar dispersion measures provide the column density, $\mathrm{DM} = \int_0^D n_e dl$, of the free electrons along the line of sight to the pulsar. H$\alpha$ emission results from recombination in ionized regions; it is proportional to the square of the electron density along the entire line of sight: $\mathrm{EM} = \int_0^\infty n_e^2 dl$.

\section{H$\alpha$ data}

\begin{figure}[tb]
\plotone{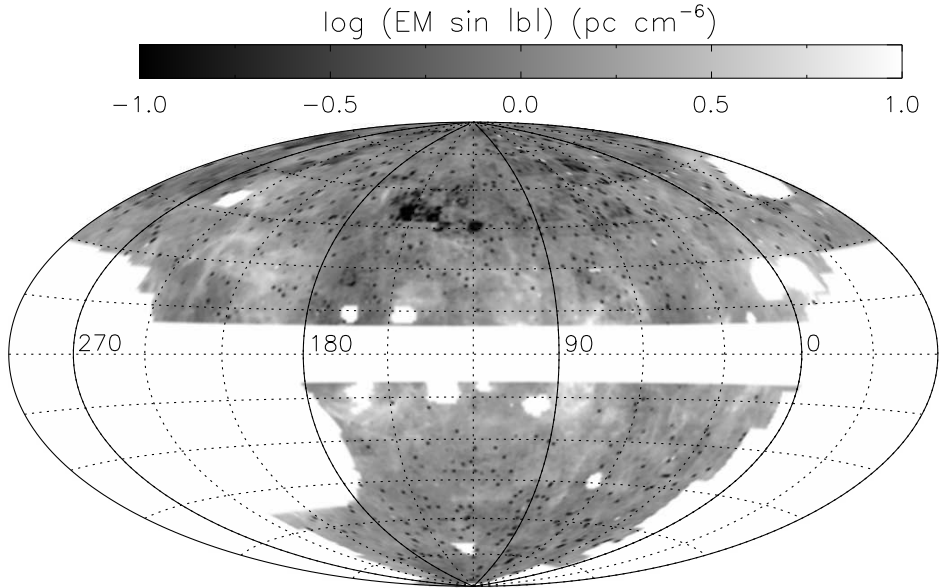}
\caption{Map of $\textrm{EM} \sin |b|$ from the WHAM Northern Sky Survey for Galactic latitude $|b| > 10\deg$ with sightlines that intersect known classical H~II regions eliminated. Dashed lines represent $30\deg$ increments in longitude and $15\deg$ increments in latitude.}
\label{fig:whammap}
\end{figure}

The Wisconsin H-Alpha Mapper (WHAM) Northern Sky Survey \citep[WHAM-NSS;][]{h03} mapped the H$\alpha$ ($\lambda 6563$) optical emission line in the northern sky with a spectral resolution of $12 \textrm{ km s}^{-1}$ and an angular resolution of $1 \deg$. We use a restricted set of the WHAM-NSS northern sky map integrated over all velocities. We have removed lines of sight that intersect classical H~II regions, leaving a sample of only the diffuse WIM (Figure~\ref{fig:whammap}).

\section{Distribution of H$\alpha$ emission}

\begin{figure}[tb]
\plotone{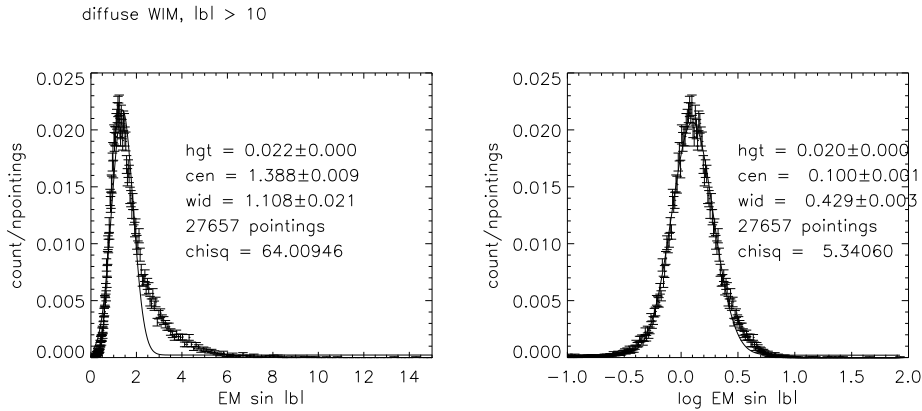}
\caption{Histograms of $\textrm{EM} \sin |b|$ ({\em left}) and $\log \textrm{EM} \sin |b|$ ({\em right}) for diffuse WIM sightlines in the WHAM-NSS survey (see text). A Gaussian fits the peak of the distribution in linear space well (dashed line), but the enhanced tail at $\textrm{EM} \sin |b| > 2 \textrm{ pc cm}^{-6}$ is typical of a lognormal distribution (dashed line, {\em right}).}
\label{fig:wim_hists}
\end{figure}

A histogram of $\textrm{EM} \sin |b|$, the component of emission measure perpendicular to the Galactic plane, for $|b|Ê>Ê10\deg$ is shown in Figure~\ref{fig:wim_hists}. We fit a normal distribution to the data (left panel of Figure~\ref{fig:wim_hists}). The sharp peak indicates that the emission measure along most sightlines is dominated by plasma in a disk-like distribution having a mean value of $\textrm{EM } \sin |b|$ of $1.1 \pm 0.5 \textrm{ pc cm}^{-6}$. There are considerable enhancements above a normal distribution at high $\textrm{EM} \sin |b|$ ($\ge 2 \textrm{ pc cm}^{-6}$), indicating the presence of high-$\textrm{EM} \sin |b|$ regions along about $1/5$ of sightlines.

The entire diffuse WIM, including the enhanced high-EM tail, is fit well by a lognormal distribution in $\textrm{EM} \sin |b|$ or, equivalently, a Gaussian distribution in $\log \textrm{EM} \sin |b|$ (right panel of Figure~\ref{fig:wim_hists}). The classical H II regions (not shown) are also well fit by a lognormal distribution with different parameters. A lognormal distribution is the expected probability density function in density for a plasma with a density structure established by isothermal turbulence \citep[e.\ g.,][]{vp99}. We are currently examining how a lognormal distribution in EM can be used to diagnose the turbulence in the WIM.

\section{Density and occupation length of the diffuse gas}

Comparisons of pulsar dispersion measures and H$\alpha$ emission measures along the same sightline provide information about the distribution of the gas which contributes to the EM and DM. There are $194$ pulsars along diffuse WIM sightlines in the WHAM-NSS region \citep{m05}. For high-EM sightlines ($\textrm{EM} \sin |b| > 2 \textrm{ pc cm}^{-6}$), the mean dispersion measure is $\langle \textrm{DM } \sin |b| \rangle = 14.8 \pm 0.9 \textrm{ pc cm}^{-3}$, whereas for low-EM sightlines ($\textrm{EM} \sin |b| < 2 \textrm{ pc cm}^{-6}$), the mean value is $\langle \textrm{DM } \sin |b| \rangle = 13.5 \pm 0.5 \textrm{ pc cm}^{-3}$. Thus, the fluctuations in density that dominate the emission measure along the high-EM sightlines contribute little to the total H$^+$ column density.

\begin{figure}[tb]
\plotone{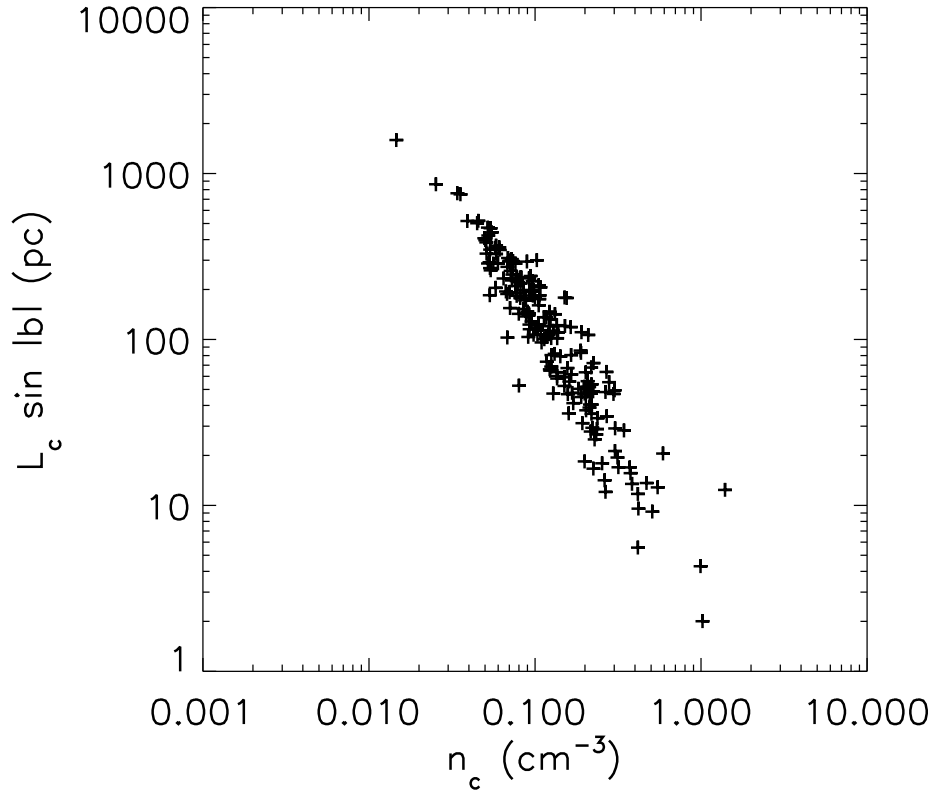}
\caption{Occupation length $L_c$ versus characteristic density $n_c$ for gas along $194$ pulsar sightlines which are in the WHAM-NSS survey area and not towards classical H~II regions.}
\label{fig:lc_nc}
\end{figure}

The characteristic electron density $n_c$ and occupation length $L_c$ of an ionized region can be estimated from
$n_c = \mathrm{EM} / \mathrm{DM}$ and
$L_c = (\mathrm{DM})^2 / \mathrm{EM}$
\citep{r91}. Figure~\ref{fig:lc_nc} shows the relationship between the density and size of ionzied regions. The gas along sightlines with densities $n_c \sim 0.1 \textrm{ cm}^{-3}$ typically occupies a few hundred parsecs along the line of sight, whereas the high-EM gas has densities approaching $1 \textrm{ cm}^{-3}$ that occupy only a few parsecs along the line of sight.

The local electron density and occupation length of the diffuse H$^+$ can be determined toward pulsars in globular clusters more than 3 kpc above the Galactic midplane, where DM samples the total H$^+$ column density through the disk. Eight such globular clusters are known to contain pulsars, but two are outside the WHAM-NSS survey region. Data for these globular clusters are shown in Table~\ref{tbl:gc_data}.

\begin{table}[bt]
\centering
\begin{tabular}{l r@{.}l r@{.}l r@{.}l r }
\hline
\hline
\multicolumn{1}{c}{Globular} & \multicolumn{2}{c}{$\textrm{DM} \sin |b|$} & \multicolumn{2}{c}{$\textrm{EM} \sin |b|$} & 
	\multicolumn{2}{c}{$n_c$} & \multicolumn{1}{c}{$L_c \sin |b|$} \\
\multicolumn{1}{c}{Cluster} & \multicolumn{2}{c}{($\textrm{pc cm}^{-3}$)} & \multicolumn{2}{c}{$(\textrm{pc cm}^{-6})$} &
	\multicolumn{2}{c}{($\textrm{cm}^{-3}$)} & \multicolumn{1}{c}{(pc)} \\
\hline
 M3 &   25&888 &  0&88	& 0&034		& 760  \\
 M5 &   21&796 &  1&3	& 0&061		& 360  \\
M13 &   19&577 &  1&0	& 0&051		& 380  \\
M15 &   30&684 &  3&2	& 0&10		& 300  \\
M30 &   18&283 &  1&7	& 0&094		& 190  \\
M53 &   23&621 &  1&1	& 0&046		& 520  \\
\hline
\end{tabular}
\caption{Properties of lines of sight towards $|z| > 3$ kpc globular clusters with known pulsars. EM data from WHAM-NSS; DM data from the ATNF pulsar catalog \citep{m05}. Sightlines towards M3 and M5 are contaminated by bright stars, so we used an average of nearby sightlines to determine the EM.}
\label{tbl:gc_data}
\end{table}

\section{Summary}

We are investigating the density and distribution of the diffuse plasma in the warm ionized medium. The H$\alpha$ emission from the WIM follows a lognormal distribution in $\textrm{EM} \sin |b|$. A lognormal distribution in density is typical of isothermal turbulence, but its significance for integrated density (EM and DM) is not yet clear. Pulsar dispersion measure data indicate that localized (a few parsecs in length), high-density ($> 0.1 \textrm{ cm}^{-3}$) regions dominate the emission measure at the high end of the EM distribution while contributing little to the total DM.


\begin{thebibliography}{}
\expandafter\ifx\csname natexlab\endcsname\relax\def\natexlab#1{#1}\fi

\bibitem[{Haffner {et~al.}(2003)Haffner, Reynolds, Tufte, Madsen, Jaehnig, \&
  Percival}]{h03}
Haffner, L.~M., Reynolds, R.~J., Tufte, S.~L., Madsen, G.~J., Jaehnig, K.~P.,
  \& Percival, J.~W. 2003, \apjs, 149, 405

\bibitem[{Manchester {et~al.}(2005)Manchester, Hobbs, Teoh, \& Hobbs}]{m05}
Manchester, R.~N., Hobbs, G.~B., Teoh, A., \& Hobbs, M. 2005, \aj, 129, 1993

\bibitem[{Reynolds(1991)}]{r91}
Reynolds, R.~J. 1991, \apjl, 372, L17

\bibitem[{{V\'{a}zquez-Semadeni} \& Passot(1999)}]{vp99}
{V\'{a}zquez-Semadeni}, E. \& Passot, T. 1999, in Interstellar Turbulence, ed.
  J.~Franco \& A.~{Carrami\~{n}ana} (Cambridge University Press), 223

\end{thebibliography}
\end{document}